\documentclass[12pt,a4paper,showpacs]{article}

\usepackage{amsmath,amssymb}
\usepackage{fleqn,epsfig}

\newcommand{\bc}{\begin{center}}
\newcommand{\ec}{\end{center}}
\newcommand{\be}{\begin{equation}}
\newcommand{\ee}{\end{equation}}
\newcommand{\ba}{\begin{array}}
\newcommand{\ea}{\end{array}}
\newcommand{\bea}{\begin{eqnarray}}
\newcommand{\eea}{\end{eqnarray}}
\newcommand{\bt}{\begin{tabular}}
\newcommand{\et}{\end{tabular}}

\newcommand{\bsl}{\boldsymbol}
\newcommand{\ov}{\overline}
\begin{document}
\begin{center}
{\Large\bf \boldmath Diquark and triquark correlations in the
deconfined phase of QCD}

\vspace*{6mm}
{I.~M.~Narodetskii, Yu.~A.~Simonov, and A.~I.~Veselov }\\[5mm]

Institute of Theoretical and Experimental Physics, Moscow
117218\\[5mm]
\end{center}
\begin{abstract}
We use the non-perturbative $Q\ov{Q}$ potential at finite
temperatures derived in the Field Correlator Method to obtain
binding energies for the lowest eigenstates in the $Q\ov{Q}$ and
$QQQ$ systems ($Q=c,b$). The three--quark problem is solved by the
hyperspherical method. The solution provides an estimate of the
melting temperature and the radii for the different diquark and
triquark bound states. In particular we find that $J/\psi$ and
$ccc$ ground states survive up to $T\sim1.3\,T_c$, where $T_c$ is
the critical temperature, while the corresponding bottomonium
states survive even up to higher temperature, $T\sim2.3T_c$.
\vspace{4mm}

\end{abstract}\vspace{2mm}

\section{Introduction} Diquark and triquark correlations in the quark--gluon
plasma (QGP) are important for understanding the dynamics of
heavy--ion collisions, processes in the early Universe and
possibly the cores of the neutron stars \cite{MN:2006}. Whether
hadrons survive in the deconfined QGP is one of the key questions
in QCD. In particular, the behavior of charmonia across the
deconfinement transition has been suggested~\footnote{~This effect
was first first investigated in a phenomenological potential model
taking into account the Debye screening \cite{MS:1986}.} as hard
probes of the QGP. Above $T_c$, the critical temperature, most
part of static $Q\ov{Q}$ interactions disappear that eventually
implies a dissolution of heavy quarkonia bound states into the
continuum. A suppression of heavy quarkonia production in heavy
ion collisions is usually considered as an observed signal of
deconfinement.

On quite general grounds it is expected that the $Q\ov{Q}$
interactions get modified by temperature. As it was first
mentioned in Ref. \cite{Si:1991} an important part of the static
interaction survives at the temperature above transition and can
support the bound $Q\ov{Q}$ states  even after the deconfinement
transition. Later existing of charmonia bound states at
$T\,\geq\,T_c$ was confirmed on the lattice \cite{ref:lattice}.

Most recently, in line with the study of Ref. \cite{Si:1991}, a
new approach called the Field Correlator Method (FCM), was
proposed to study dynamics of QGP, where the main emphasis was
done on the nonperturbative vaccuum fields
\cite{ST},\cite{Si:2008},\cite{NST:2009}. Let us summarize the
basic formulation of the FCM as applied to finite T. The approach
is based on the study of the quadratic field correlators
$<tr\,F_{\mu\nu}(x)\Phi(x,0)F_{\lambda\sigma}(0)>$ ($x$ is
Euclidian), where $\Phi(x,0)$ is the parallel transporter
necessary to maintain gauge invariance. The Gaussian correlator is
expressed in terms of two scalar functions, $D(x)$ and $D_1(x)$,
which define the static potential between heavy quarks at
$T\,=\,0$, the confinement part of the potential, $\sigma\,r$, of
is expressed only in terms of
$D(x)$:\be\sigma\,=\,2\,\int\limits_0^{\infty}\,d\lambda\,\int\limits_0^{\infty}\,d\tau\,D(\sqrt{\lambda^2+\tau^2})
.\ee At $T\,\geq\,T_c$ one should distinguish between electric and
magnetic correlators $D^E(x)$, $D^H(x)$, $D^E_1(x)$, and
$D^H_1(x)$, and, correspondingly, between $\sigma^E$ and
$\sigma^H$. It was argued in \cite{Si:1992} and later confirmed on
the lattice \cite{DiGiacomo} that above the deconfinement region
$D^E(x)$ and, correspondingly, $\sigma^E$ vanish, while the
colorelectric correlator $D_1^E(x)$ and colormagnetic correlators
$D^H(x)$ and $D^H_1(x)$ should stay unchanged at least up to
$T\sim 2\,T_c$. However, the correlators $D^H(x)$ and $D^H_1(x)$
do not produce static quark--antiquark potentials, they only
define the spatial string tension $\sigma_s\,=\,\sigma^H$ and the
Debye mass $m_D\propto\sqrt{\sigma_s}$ that grows with the
temperature in the dimensionally reduced limit \cite{AS:2006}.

The main source of the quark--antiquark static interaction at
$T\,\geq\,T_c$ originates from the correlator $D^{E}_1(x)$
\cite{DMSV:2007}\be\label{eq:potential}
V_{Q\ov{Q}}^{np}(r,T)\,=\,\int\limits_0^{1/T}d\nu(1-\nu
T)\int\limits_0^r \lambda d\lambda\,
D_1^{E(np)}(\sqrt{\lambda^2+\nu^2}),\ee that is responsible for
bound states of quarks and gluons in the QGP .

The purpose of this Letter is to report our latest results of low
lying charmonium and bottomonium as well as $ccc$, $bbb$ bound
states above $T_c$, obtained from the use of this approach. The
paper is organized as follows. In  section \ref{section:HQP} we
introduce the static $Q\ov{Q}$ potential. In section
\ref{section:quarkonia} the numerical results for the ground
states of $J/\psi$ and $\Upsilon$ are presented. In section
\ref{section:QQQ} we evaluate the masses of the  three-quark $ccc$
and $bbb$ baryons. Our conclusions are given in section
\ref{section:conclusions}
\section{The heavy quark-antiquark potential at $T\geq T_c$} \label{section:HQP} In the framework of the FCM, the
finite temperature behavior of the static $Q\bar Q$ potential
$V_{Q\overline Q}(r,T)$ at $T\geq T_c$ is given by
\be\label{eq:VQQ}
 V_{Q\bar Q}(r,T) = V_{Q\bar Q}^{pert}(r,
T) + V_{Q\bar Q}^{np}(r,T).\ee In Eq. (\ref{eq:VQQ}) $ V_{Q\bar
Q}^{pert}(r, T)$ describes the interaction at short distances
whereas $ V_{Q\bar Q}^{np}(r,T)$ is the long-distance potential.
The short distance interaction is represented by the perutrbative
one--gluon exchange potential that undergoes a Debye screening by
the color charges of the QGP \be \label{eq:Coulomb} V_{Q\bar
Q}^{\rm pert}(r,
T)\,=\,-\,C_F\,\frac{\alpha_s}{r}\,\exp(-m_D\,r),\ee where
$C_F=4/3$ is the color factor, $m_D$ is the inverse of the Debye
screening radius. This form of the modified Coulomb potential has
been used in earlier works (see {\it e.g.} Ref. \cite{Digal:2001})
to specify the in-medium potential between heavy quarks and
determine the dissociation points of different quarkonium states.
The Debye mass $m_D$ in Eq. (\ref{eq:Coulomb}) can be written as $
m_D\,=\,2.06\,\sqrt{\sigma_s(T)}$, where $\sigma_s(T)$ is the
spatial string tension due to  chromomagnetic confinement. In what
follows we use the results of Ref. \cite{Agasian:2003} where the
quantity $\sqrt{\sigma_s(T)}$ has been computed nonperturbatively
up to two loops in the deconfined phase of QCD. For $n_f=0$ the
Debye mass varies between 0.8 GeV and 1.4 GeV, when $T$ varies
between $T_c=275$ MeV and $2\,T_c$.

\begin{figure}
\begin{center}
\epsfxsize=7.5cm \epsfysize=7.5cm \epsfbox{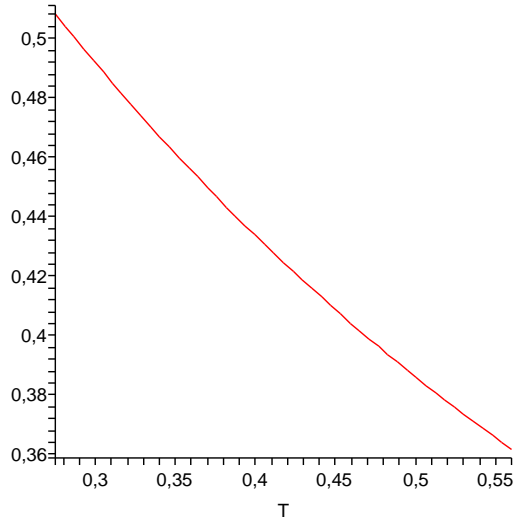}
\end{center}
\caption{The dependence of the $V_{Q\ov{Q}}^{np}(\infty,T)$ on the
temperature $T$ given by Eq. (\ref{eq:alpha}).}
\label{fig:vqq(infinity)}
\end{figure}

The long-distance interaction $V^{np}(r,T)$ requires theoretical
assumptions about its shape. We follow Ref. \cite{Si:2005}, where
the long distance nonperturbative $Q\ov{Q}$ potential was
 derived analytically from the analysis of the
non--perturbative part of the correlator function $D^E_1(x)$:
\be\label{eq:asymptotics}
D^{E\,(np)}_1(x)\,=\,{B}\exp(-M_0\,|x|)/|x|\ee In Eq.
(\ref{eq:asymptotics}) the coefficient $B$ must be considered as
functions of the physical temperature. In the confinement region
\be B=2C_F\alpha_s^f\sigma_{adj}M_0\,=\,6\alpha_s^f\sigma_fM_0,\ee
$\alpha_s^f$ being the freezing value of the strong coupling
constant in the confinement region, $\sigma_f$ is zero temperature
string tension, and $M_0$ has the meaning of the lowest gluelump
mass
 \cite{S:2001}. Taking $\alpha_s^f=0.6$, $\sigma_f=0.18$ GeV$^2$,
$M_0=0.9$ GeV, we get $B=0.583$ GeV$^3$. Above the critical
temperature one substitutes $B\,\rightarrow\,\xi(T)B$, where the
$T$--dependent constant  \be
\xi(T)\,=\,\left(1-0.36\,\frac{M_0}{B}\,\frac{T-T_c}{T_c}\right)\ee
arises from the comparison with lattice data \cite{DMSV:2007}. As
the result one obtains
 \be\label{eq:VQQnp}
V_{Q\ov{Q}}^{np}(r,T)=\xi(T)\,
\left(V^{np}(\infty,T)-V^{np}(r)\right),\ee
where the continuum threshold (a constant shift in the potential)
$V^{np}(\infty,T)$ in Eq. (\ref{eq:VQQnp}) is \be\label{eq:alpha}
V^{np}(\infty,T)\,=\,\frac{B}{M_0^2}\,\left[1\,-\,\frac{T}{M_0}\,\left(1\,-\,\exp(-\,\frac{M_0}{T})\right)\right],\ee
and \be\label{eq:u(r)} V^{np}(r)=\frac{B}{M_0^2}\,\,x\,K_1(x)+
{\cal{O}}\left(\frac{T}{M_0}\right).\ee In Eq. (\ref{eq:u(r)})
  $K_1(x)$ is the modified Bessel
 function and $x=M_0r$.
At $T=T_c$ one obtains $V_{Q\ov{Q}}^{np}(\infty,T_c)\approx 0.5$
GeV that agrees with estimate obtained in Ref. \cite{ST} from
lattice data. The large positive value of the continuum threshold
is a consequence of non perturbative vaccuum fields and can not be
explained in perturbation theory. The behavior of the potential
$V_{Q\overline{Q}}(\infty,T)$ is shown in Fig.
\ref{fig:vqq(infinity)}

\section{Quarkonia at finite temperature}
\label{section:quarkonia} Having specified the static $Q\bar Q$
potential, we can now exploit the relativistic Hamiltonian
technique, developed in \cite{DKS:1994} to calculate the masses of
the $Q\bar Q$ states as a function of the temperature.  This
technique does not take into account chiral degrees of freedom and
is applicable when spin-dependent interaction can be treated as
perturbation. Therefore below  consider heavy quarkonia  and
baryons. The masses of heavy quarkonia are defined as \be M_{Q\bar
Q}\,=\,\frac{m_Q^2}{\mu_{Q}}\,+\,\mu_Q\,+\,E_0(m_Q,\mu_Q),\ee
where $E_0(m_Q,\mu_Q)$ is an eigenvalue of the Hamiltonian
$H\,=\,H_0\,+\,V_{Q\bar Q}$, where $V_{Q\bar Q}$ is given by Eq.
(\ref{eq:VQQ}), $m_Q$ are the bare quark masses, and $\mu_Q$ are
the constant auxiliary fields (AF) that were introduced  to treat
the kinematics of the relativistic particles. These parameters
have to be found from the variational condition \be\frac{\partial
M_{Q\bar Q}}{\partial \mu_Q}\,=\,0\ee The eigenvalue problem is
solved for each set of $\mu_Q$; then one has to minimize $M_{Q\bar
Q}$ with respect to $\mu_Q$. Such an approach allows for a very
transparent interpretation of AF: starting from bare quark masses
$m_Q$, we naturally arrive at the dynamical masses $\mu_Q$ that
appear due to the interaction. The AF are treated as c-number
variational parameters. The bound state exists if
$E_0(m_Q,\mu_Q)\,\leq\,V^{np}(\infty,T)$.

To find out whether the non perturbative interaction can support
the bound states at $T\,\sim\,T_c$ we use the Bargmann condition
\be n\leq I\,=\,\mu_Q
\int\limits_0^{\infty}\,|V^{np}(r)\,+\,V_{Q\bar Q}^{pert}(r,
T)|\,r\,dr\,=\,\mu_Q\left(\frac{4\alpha_s}{3\,m_d}+\frac{B}{M_0^4}\right),\ee
where $n$ is the number of the $S$--wave bound states. Since
$m_d\sim 1$ GeV, $B/M_0^2\sim 0.5$ GeV, and $M_0\sim 1$ GeV, we
conclude that to support at least one bound state one needs
$\mu_Q\geq 1$ GeV, {\it i.e.} there is no bound states of light
quarks, but $c\ov{c}$ and especially $b\ov{b}$ binding is
possible. Moreover, the bottomonium spectrum should display a much
larger number of bound states above $T_c$.

The solutions for the binding energy for the $1S$ states of
charmonium and bottomonium are shown in Tables \ref{tab:cc},
\ref{tab:bb}. In these Tables we present the constituent quark
masses $\mu_Q$ for $c\ov{c}$ and $b\ov{b}$~\footnote{These masses
are computed solely in terms of the bare quark masses $m_c$ and
$m_b$, respectively. Note that, as in the confinement region
\cite{NTV}, the constituent masses $\mu_Q$ only slightly exceed
bare quark masses $m_Q$ that reflect smallness of the kinetic
energies of heavy quarks.}, the differences
$\varepsilon_Q=E_0\,-\,V_{Q\overline{Q}}(\infty)$, the mean
squared radii $\sqrt{\,\overline{r^2}\,}$, and the masses
$M_{Q\ov{Q}}$ for the $c\ov{c}$ and $b\ov{b}$ mesons. We employ
$m_c=1.4$ GeV, $m_b=4.8$ GeV, and $\alpha_s\,=\,0.35$. As
expected, we obtain the weakly bound state at $T=T_c$ that
disappears  at $T\sim 1.3\,T_c$. The charmonium masses lie in the
interval 3.2 - 3.3 GeV, that agrees with the results of Ref.
\cite{DMSV:2007}. Note that immediately above $T_c$ the mass of
the  $c\ov{c}$ state is about 0.2 GeV higher than that of
$J/\psi$.
\begin{table}\caption{The $1S\,\,\,J/\psi$ state above the deconfinement region. ~$M_0\,=\,0.9$ GeV,~$B\,=\,0.583$
GeV$^3$,~$m_c\,=\,1.4$ GeV, $\alpha_s\,=\,0.35$. Masses and
energies are given in units of GeV, the mean squared radius
$\sqrt{<r^2>}$ in units of GeV$^{-1}$.}\vspace{6mm}

\centering
\begin{tabular}{c|cccc}\hline\hline\\
$T/T_c$&$\mu_c$&$\varepsilon_c$&$\sqrt{<r^2>}$&$M_{c\overline{c}}$\\\\\hline\\
1~~&~~1.451&-\,0.019&7.53&3.291\\
1.3~~&~~1.419&+\,0.006&10.50&3.186\\\\

\hline\hline
\end{tabular}
\vspace{5mm}

 \label{tab:cc}
\end{table}

\begin{table}
\caption{The $1S\,\,\,\Upsilon$ state above the deconfinement
region.~The notations are the same as in Table
\ref{tab:cc},~$m_b\,=\,4.8$ GeV}.\vspace{6mm}

\centering
\begin{tabular}{c|cccc}\hline\hline\\
$T/T_c$&$\mu_b$&$E_0\,-\,V_{Q\ov{Q}}(\infty)$&$\sqrt{<{r^2}>}$&$M_{b\ov{b}}$\\\\\hline\\
1~~&~~4.984&-\,0.300&1.27&9.815\\
1.3~~&~~4.950&-\,0.183&1.55&9.802\\
1.6~~&~~4.915&-\,0.095&2.06&9.783\\
2.0~~&~~4.863&-\,0.021&4.25&9.742\\
2.2~~&~~4.832&-\,0.003&7.62&9.712\\
2.3~~&~~4.818&+\,0.001&9.52&9.694\\

\\ \hline\hline
\end{tabular}
\label{tab:bb}\end{table}
 \vspace{5mm}

 As expected, the $b\overline{b}$ bound states remain intact up to the
 larger
temperatures, $T\,\sim\, 2.3\,T_c$, see Table \ref{tab:bb}. The
masses of the L = 0 bottomonium  lie in the interval 9.7--9.8 GeV,
about 0.2--0.3 GeV higher than 9.460 GeV, the mass of
$\Upsilon(1S)$.

The results for $c\ov{c}$ and $b\ov{b}$ bound states found from
the FCM can be compared to the calculations based on
phenomenological $Q\ov{Q}$ potentials identified with the free
energy measured on the lattice \cite{Blaschke:2005},
\cite{Alberico}. Our results for $1S(J/\psi)$ are qualitatively
agree with those of Refs. \cite{Blaschke:2005}, \cite{Alberico}
while our melting temperature for $1S(\Upsilon)$ is much smaller
than $T\sim (4-6)\,T_c$ found in Ref. \cite{Alberico}.

 \vspace{2mm}

\section{$QQQ$ baryons at $T\geq T_c$}\label{section:QQQ}
The three quark potential is given by \be\label{eq:QQQ}
V_{QQQ}\,=\,\frac{1}{2}\,\sum_{i<j}\,V_{Q{\bar Q}}(r_{ij},T),\ee
where $\frac{1}{2}$ is the color factor.
We solve the three quark Schr\"{o}dinger equation using the
hyperspherical method. The wave function of a $QQQ$ baryon depends
on the three-body Jacobi coordinates \be\label{eq:Jacobi}
\bsl{\rho}_{ij}=\sqrt{\frac{\mu_{Q}}{2}}\,(\bsl{r}_i-\bsl{r}_j),\,\,\,\,\,\,\,\,\,\,\,
\bsl{\lambda}_{ij}=\sqrt{\frac{2}{3}\,\mu_Q}
\left(\frac{\bsl{r}_i+\bsl{r}_j}{2}-\bsl{r}_k\right),\ee ($i,j,k$
cyclic). There are three equivalent ways of introducing the Jacobi
coordinates, which are related to each other by linear
transformations with the Jacobian equal to unity. In what follows
we omit the indices $i$ and $j$.

In terms of the Jacobi coordinates the three--quark kinetic energy
operator $H_0$ is written as \be \label{H_0_jacobi} H_0=
-\frac{1}{2} \left(\frac{\partial^2}{\partial\bsl{\rho}^2}
+\frac{\partial^2}{\partial\bsl{\lambda}^2}\right)\,=\,
-\,\,\frac{1}{2}\left( \frac{\partial^2}{\partial
R^2}+\frac{5}{R}\frac{\partial}{\partial R}+
\frac{\bsl{L}^2(\Omega)}{R^2}\right), \ee where $R$ is the
six-dimensional hyperradius that is invariant under quark
permutations, \be R^2\,=\,\bsl{\rho}^2+\bsl{\lambda}^2\,=\,
\frac{\mu_Q}{3}\,\left(r_{12}^2\,+\,r_{23}^2\,+\,r_{31}^2\right),\ee
\be\rho\,=\,R\,\sin\theta,\,\,\,\,\,
\lambda\,=\,R\,\cos\theta,\,\,\,\, 0 \le \theta \le \pi/2,\ee
$\Omega$ denotes five residuary angular coordinates, and
$\bsl{L}^2(\Omega)$ is an angular operator
 \be{\bf
L}^2\,=\,\frac{\partial^2}{\partial
\theta^2}\,+\,4\cot\theta\,\frac{\partial}{\partial
\theta}-\frac{{\bf l}_{\rho}^2}{\sin^2\theta}\,-\,\frac{{\bf
l}_{\lambda}^2}{\cos^2\theta},\ee
whose eigenfunctions (the hyperspherical harmonics) satisfy
\begin{equation}
\label{eq: eigenfunctions} {\bf L}^2(\Omega)\,Y_{[K]}(\theta,{\bf
n}_{\rho},{\bf n}_{\lambda})\,=\,-K(K+4)Y_{[K]}(\theta,{\bf
n}_{\rho},{\bf n}_{\lambda}),
\end{equation}
with $K$ being the grand orbital momentum.

The wave function $\psi(\bsl{\rho},\bsl{\lambda})$ is written in a
symbolical shorthand as
\be\psi(\bsl{\rho},\bsl{\lambda})=\sum\limits_{[K]}\psi_{[K]}(R)Y_{[K]}(\Omega),\label{eq:ss}\ee
where the set $[K]$ is defined by  the orbital momentum of the
state and the symmetry properties. We truncate this set using the
hypercentral approximation $K\,=\,K_{\rm min}\,=\,0$. Introducing
the reduced function $u(R)$
\be\Psi(\bsl{\rho},\bsl{\lambda},T)\,=\,\frac{1}{\sqrt{\pi^3}}\,\frac{u(R,T)}{R^{5/2}},
\ee
and averaging the three--quark potential (\ref{eq:QQQ}) over the
six-dimensional sphere one obtains the one-dimensional
Schr\"odinger equation for $u(R,T)$

 \be\label{eq:se}\frac{d^2
u(R,T)}{dR^2}\,+\,2\left[E_0\,-\,\frac{15}{8\,R^2}\,-
\,\frac{3}{2}\,\xi(T)\,\left({\cal V}^{\rm pert}(R,T)\,+\,{\cal
V}^{np}(R,T)\right)\right]\,u(R,T)\,=\,0,\ee where \be {\cal
V}^{\rm pert}(R,T)\,=\,
-\frac{4}{3}\,\alpha_s\,\frac{a_C(R)}{R},\ee \be a_C(R)\,=\,\frac{
16}{\pi}\,\sqrt{\frac{\mu_{Q}}{2}}\,\int\limits_0^{\pi/2}
\exp\left(\,\frac{-m_D\,R\,\sin\theta}{\sqrt{\mu_{Q}/2}}\right)\,\sin\theta\cos^2\theta\,d\theta,\ee
and \be {\cal V}^{\rm np}(R,T)=V^{np}(\infty,T)-\frac{16B}{\pi
M_0}\,\sqrt{\frac{2}{\mu_Q}}\,\left(\int\limits_0^{\pi/2}K_1\left(\frac{M_0R
\sin\theta}{\sqrt{\mu_{Q}/2}}\right)\sin^3\theta\cos^2\theta
d\theta\right)R.\ee
The temperature dependent mass of the colorless $QQQ$ states is
defined as \be M_{QQ
Q}\,=\,\frac{3}{2}\frac{m_Q^2}{\mu_{Q}}\,+\,\frac{3}{2}\,\mu_Q\,+\,E_0(m_Q,\mu_Q).\ee
The bound $QQQ$ state exists if
$E_0(m_Q,\mu_Q)\,\leq\,V_{QQQ}(\infty,T)$, where \be {\cal
V}_{QQQ}(\infty,T)\,=\,\frac{3}{2}\,\,V^{np}(\infty,T).\ee In
Tables \ref{tab:ccc} and \ref{tab:bbb} we show the masses of the
ground $ccc$ and $bbb$ states as a function of the temperature.
\begin{table}
\caption{The ground $ccc$ state as a function of the temperature
above the deconfinement region.~The notations are the same as in
Table \ref{tab:cc}}\vspace{6mm}

\centering
\begin{tabular}{c|cccc}\hline\hline\\
$T/T_c$&$\mu_c$&$E_0\,-\,\cal{V}_{QQQ}(\infty)$&$\sqrt{<R^2>}$&$M_{ccc}$\\\\\hline\\
1~~&~~1.474&-\,0.046&6.60&4.922\\
1.3~~&~~1.434&-\,0.002&9.96&4.769\\

\\ \hline\hline
\end{tabular}
 \vspace{3mm}

\label{tab:ccc}
\end{table}

\begin{table}\caption{The ground $bbb$ state
as a function of the temperature above the deconfinement
region.~The notations are the same as in Table
\ref{tab:cc}}\label{tab:bbb}\vspace{6mm}

\centering
\begin{tabular}{c|cccc}\hline\hline\\
$T/T_c$&$\mu_b$&$E_0\,-\,\cal{V}_{QQQ}(\infty)$&$\sqrt{<R^2>}$&$M_{bbb}$\\\\\hline\\
1~~&~~4.977&-\,0.506&3.12&14.665\\
1.3~~&~~4.927&-\,0.314&3.86&14.661\\
1.6~~&~~4.885&-\,0.171&5.04&14.645\\
2.0~~&~~4.840&-\,0.046&8.21&14.598\\
2.2~~&~~4.822&-\,0.011&10.59&14.562\\
2.3~~&~~4.815&+\,0.001&11.62&14.541\\

\\ \hline\hline
\end{tabular}
 \vspace{3mm}
\end{table}
\vspace{1mm} \section{Conclusions}\label{section:conclusions}
Based on the non-perturbative dynamics driven by the field
correlators at finite temperatures
we have calculated binding energies for the lowest eigenstates in
the $c\ov{c}$, $ccc$, $b\ov{b}$,and $bbb$ systems. For what
concerns the charmonium we find that $J/\psi$ (and $ccc$ ground
states) survive up to $T\,\sim\,1.3\,T_c$. On the other hand, the
$b\ov{b}$ and $bbb$ states survive even up to higher temperature,
$T\,\sim\,2.3\,T_c$. This suggests that the systems are  strongly
interacting above $T_c$.

\vspace{1mm}

This work was supported in part by RFBR Grants
$\#\,\,$08-02-00657, $\#\,\,$08-02-00677, $\#\,\,$09-02-00629 and
by the grant for scientific schools $\#\,\,$NSh.4961.2008.2.



\end{document}